\documentstyle{mn}

\newif\ifAMStwofonts
\newcommand{\xmm}{{\it XMM-Newton}}

\newcommand{\pn}{{\it pn}}
\newcommand{\mos}{{\it MOS}}

\newcommand{\msol}{$M_{\odot}$}

\newcommand{\ha}{H$\alpha$}
\newcommand{\hb}{H$\beta$}
\newcommand{\hg}{H$\gamma$}

\newcommand{\sfr}{M$_\odot$yr$^{-1}$}
\ifoldfss
  \ifCUPmtlplainloaded \else
    \NewTextAlphabet{textbfit} {cmbxti10} {}
    \NewTextAlphabet{textbfss} {cmssbx10} {}
    \NewMathAlphabet{mathbfit} {cmbxti10} {} % for math mode
    \NewMathAlphabet{mathbfss} {cmssbx10} {} %  "   "    "
  \fi
  \ifAMStwofonts
    \ifCUPmtlplainloaded \else
      \NewSymbolFont{upmath} {eurm10}
      \NewSymbolFont{AMSa} {msam10}
      \NewMathSymbol{\upi}     {0}{upmath}{19}
      \NewMathSymbol{\umu}     {0}{upmath}{16}
      \NewMathSymbol{\upartial}{0}{upmath}{40}
      \NewMathSymbol{\leqslant}{3}{AMSa}{36}
      \NewMathSymbol{\geqslant}{3}{AMSa}{3E}

    \fi
  \fi
\fi % End of OFSS

\ifnfssone
  \newmathalphabet{\mathit}
  \addtoversion{normal}{\mathit}{cmr}{m}{it}
  \addtoversion{bold}{\mathit}{cmr}{bx}{it}
  \newmathalphabet{\mathbfit} % math mode version of \textbfit{..}
  \addtoversion{normal}{\mathbfit}{cmr}{bx}{it}
  \addtoversion{bold}{\mathbfit}{cmr}{bx}{it}
  \newmathalphabet{\mathbfss} % math mode version of \textbfss{..}
  \addtoversion{normal}{\mathbfss}{cmss}{bx}{n}
  \addtoversion{bold}{\mathbfss}{cmss}{bx}{n}
  \ifAMStwofonts
    \ifCUPmtlplainloaded \else
      \UseAMStwoboldmath
      \makeatletter
      \new@mathgroup\upmath@group
      \define@mathgroup\mv@normal\upmath@group{eur}{m}{n}
      \define@mathgroup\mv@bold\upmath@group{eur}{b}{n}
      \edef\UPM{\hexnumber\upmath@group}
      \new@mathgroup\amsa@group
      \define@mathgroup\mv@normal\amsa@group{msa}{m}{n}
      \define@mathgroup\mv@bold\amsa@group{msa}{m}{n}
      \edef\AMSa{\hexnumber\amsa@group}
      \makeatother
      \mathchardef\upi="0\UPM19
      \mathchardef\umu="0\UPM16
      \mathchardef\upartial="0\UPM40
      \mathchardef\leqslant="3\AMSa36
      \mathchardef\geqslant="3\AMSa3E
    \fi
  \fi
\fi % End of NFSS release 1

\ifnfsstwo
  \DeclareMathAlphabet{\mathbfit}{OT1}{cmr}{bx}{it}
  \SetMathAlphabet\mathbfit{bold}{OT1}{cmr}{bx}{it}
  \DeclareMathAlphabet{\mathbfss}{OT1}{cmss}{bx}{n}
  \SetMathAlphabet\mathbfss{bold}{OT1}{cmss}{bx}{n}
  \ifAMStwofonts
    \ifCUPmtlplainloaded \else
      \DeclareSymbolFont{UPM}{U}{eur}{m}{n}
      \SetSymbolFont{UPM}{bold}{U}{eur}{b}{n}
      \DeclareSymbolFont{AMSa}{U}{msa}{m}{n}
      \DeclareMathSymbol{\upi}{0}{UPM}{"19}
      \DeclareMathSymbol{\umu}{0}{UPM}{"16}
      \DeclareMathSymbol{\upartial}{0}{UPM}{"40}
      \DeclareMathSymbol{\leqslant}{3}{AMSa}{"36}
      \DeclareMathSymbol{\geqslant}{3}{AMSa}{"3E}
    \fi
  \fi
\fi % End of NFSS release 2

\ifCUPmtlplainloaded \else
  \ifAMStwofonts \else % If no AMS fonts
    \def\upi{\pi}
    \def\umu{\mu}
    \def\upartial{\partial}
  \fi
\fi

\title[Evolution of the X-ray Luminosity in Young HII Galaxies]
{Evolution of the X-ray Luminosity in Young HII Galaxies\thanks{Partially based on observations obtained with \xmm , 
an ESA science mission with instruments and contributions directly funded by ESA Member States and NASA.}}
\author[D. Rosa Gonz\'alez et al.]
       {D. Rosa Gonz\'alez$^1$, E. Terlevich$^1$\thanks{Visiting Fellow, IoA,
           Cambridge, UK}, 
E. Jim\'enez Bail\'on$^2$, R. Terlevich$^{1\dag}$, P. Ranalli$^3$, 
\newauthor A. Comastri$^4$, E. Laird$^5$ and K. Nandra$^5$ 
        \\
      $^1$  Instituto Nacional de Astrof\'{\i}sica Optica y Electr\'onica.
        Luis Enrique Erro No. 1. Tonantzintla, Puebla, C.P. 72840, M\'exico
        \\ 
$^2$ Instituto de Astronom\'\i a, Universidad Nacional Aut\'onoma de M\'exico,
        Apartado Postal 70-264, 04510-M\'exico DF, M\'exico
\\ 
$^3$ Universit\'a di Bologna - Dipartimento di Astronomia - via Ranzani 1 - 40127
Bologna, Italy
\\
$^4$ INAF - Osservatorio Astronomico di Bologna, via Ranzani 1, 40127 Bologna, Italy
\\
$^5$ Astrophysics Group, Imperial College London, Blackett Laboratory, Prince Consort Road, London SW7 2AZ
}
\date{Accepted 2009 June 23.
      Received 2009 June 23;
      in original form 2008 August 11}

\pagerange{\pageref{firstpage}--\pageref{lastpage}}
\pubyear{2009}

\begin{document}

\maketitle

\label{firstpage}

\begin{abstract}
In an effort to understand the correlation between X-ray emission and 
present star formation rate (SFR), we obtained XMM-Newton data to estimate the X-ray 
luminosities of a sample of actively starforming HII galaxies.
 The obtained  X-ray luminosities are compared to 
other well known tracers of star formation activity 
such as the far infrared and the ultraviolet luminosities. We also compare 
the obtained results with empirical laws from the literature
and  with recently published analysis applying synthesis models.
We use the time delay between the formation of the stellar 
cluster and that of the first X-ray binaries, in order
to put limits on the age of a given stellar burst.
We conclude that the generation of soft X-rays, as well as the  \ha\  or 
infrared luminosities is instantaneous. The relation between the observed 
radio and hard X-ray luminosities, on the other hand, points to the existence of 
a time delay between the formation of the stellar cluster and the explosion of 
the first massive stars and the consequent formation of supernova remnants 
and high mass X-ray binaries (HMXB) 
which originate the radio and hard X-ray fluxes respectively. 
When comparing hard X-rays with a 
star formation indicator that traces the first million years
of evolution (e.g. \ha\, luminosities)  we found a deficit in the
expected X-ray luminosity. This deficit is not found when the X-ray luminosities
are compared with infrared luminosities, a star formation tracer that 
represents an average over the last 10$^8$\,years.
The results support the hypothesis that hard X-rays are originated in X-ray 
binaries which,  as  supernova remnants, have a formation  
time delay of a few mega years after the starforming burst.   
\end{abstract}

\begin{keywords}
Galaxies: evolution -- Galaxies: starburst --  
X-rays: galaxies
\end{keywords}

\section{Introduction}

X-ray emission in star forming galaxies is dominated by a combination of
high mass X-ray binaries (HMXBs), hot O-stars, young supernova remnants and 
hot plasma, all sources that are closely related to the presence of massive 
short-lived stars which directly trace the current star formation activity
~\cite{2002Persic}. Based on observations of local
galaxies and the comparison with other tracers 
of recent star formation activity, 
several authors have proposed the use of the X-ray luminosity 
as a star formation tracer 
[e.g. David, Jones, \& Forman~\cite{1992David}, Grimm, Gilfanov \& Sunyaev~\cite{2003Grimm}, 
Ranalli, Comastri \& Setti~\cite{2003Ranalli}].
The locally observed correlations appear to hold also at high-$z$ and 
an empirical L$_{\rm X}$$-$SFR relation based on observations of 
Lyman Break Galaxies (LBGs) in the 1~Ms CDF-N has been 
derived~\cite{2002Nandra}. 
However studies of LBGs suffer from 
the lack of other direct SFR tracers (e.g. \ha, FIR), and additionally, 
LBGs whose emission is dominated by starburst events are too 
faint to be directly detected. Only the {\it mean} X-ray
luminosities are attainable,  making the possible contamination by
obscured AGN difficult to quantify~\cite{2004Persic}.
Recent papers have confirmed the use of the
X-ray emission as a direct tracer of  star formation  events
[e.g. Laird et al.~\cite{2005Laird}, Rosa-Gonz{\'a}lez et
al.~\cite{2007Rosa-CDFS}]; 
in contrast, Barger, Cowie \& Wang~\cite{2007Barger} found a poor correlation
between the X-ray and radio emissions of starforming galaxies that could question the use of 
X-rays as a reliable tracer of star formation activity.
In a recent paper by
Mas-Hesse, Ot{\'{\i}}-Floranes \& Cervi{\~n}o~\cite{2008Mas}, the different 
empirical relations between X-ray luminosities and the current SFRs 
were compared against evolutionary models. They found that under realistic 
assumptions (e.g. a young burst and an efficiency  of a few percent in the 
re-processing of mechanical energy) the observed relations confirm
the use of the soft X-ray luminosity as a reliable tracer of the 
star formation activity in young systems. 

This paper discusses the X-ray emission (as detected by \xmm )
of HII galaxies selected from the Terlevich et al.~\cite{1991Terlevich} catalog as having 
intense \ha \ and \hb\ emission lines in their optical spectra.

HII galaxies are compact systems dominated by a strong and recent star 
formation burst event. The relative low mass -- which implies a low 
contamination by low mass X-ray binaries -- 
together with the absence of AGN activity  make these objects 
the best laboratories to study the relation between 
the X-ray luminosity (L$_{\rm X}$) and the current star formation rate (SFR). 
By using other tracers of  star forming activity we will discuss the validity 
of the  existing X-ray calibrations in very young systems. 
In fact, HII galaxies which are the youngest starbursts 
known in the local universe (Rosa-Gonz{\'a}lez et al.~\cite{2007Rosa} could have a deficit of X-ray 
emission due to a time lag between the formation of the massive star cluster 
and the formation of the first HMXBs.
In the following section we discuss the sample selection. In section 3, the 
data and its analysis. A comparison between the X-ray results and those 
obtained with other tracers of star formation is presented in Section 4.
Section 5 is the discussion and conclusions are given in section 6. 

Throughout this work a standard, flat $\Lambda$CDM cosmology with
$\Omega_\Lambda$= 0.7 and H$_0$ = 70 km s$^{-1}$ Mpc$^{-1}$ is assumed.

\section{Sample Selection and SFRs from Optical Emission Lines}
HII galaxies harbour intense and young star formation events revealed 
by the strong emission lines observed in the optical spectra.
From the Spectrophotometric Catalog of HII galaxies (Terlevich et al.~\cite{1991Terlevich}
we have selected those galaxies for which the corresponding SFR is greater 
than 4~\sfr. 
%%%%%%%%%%%%%%%%%%%%%%  Changed Text %%%%%%%%%%%%%%%%%%%%%%%%%%%%%%%%%%%%%%%%%%
Based on the calibration given by Grimm et al.~\cite{2003Grimm}
and the estimated SFR, we expect  to find at least 
12 HMXBs with luminosities greater than 10$^{38}$ erg\,s$^{-1}$ in each one of the
selected galaxies. Due to the galaxies low mass, the expected number 
    of LMXBs   is a minimum. The selected sample is given in Table~1.
%The low mass of the selected galaxies minimizes the contribution 
%of LMXBs that could be present in the galaxy.
%%%%%%%%%%%%%%%%%%%%%%%%%%%%%%%%%%%%%%%%%%%%%%%%%%%%%%%%%%%%%%%%%%%%%%%%%%%%%%

The luminosities of the hydrogen recombination lines are  proportional to 
the number of ionizing photons so they trace the presence of 
massive stars with lifetimes not larger than a few million years. 
We estimate the star formation rate from the reddening corrected \ha\ 
luminosities [SFR(\ha)] using the relation~\cite{1998Kennicutt},

\begin{equation}
SFR(H\alpha) (M_\odot yr^{-1}) = 7.9\times 10^{-42} L(H\alpha) (erg\,s^{-1})
\end{equation}

\ha\ and \hb\ fluxes are corrected for extinction
using the Milky Way extinction curve~\cite{1979Seaton} 
and assuming an intrinsic ratio \ha/\hb\ of 2.87.
%%%%%%%%%%%%%%%%%%%%%%  Changed Text %%%%%%%%%%%%%%%%%%%%%%%%%%%%%%%%%%%%%%%%%%
The ionizing massive young clusters show in their spectra strong   
absorptions in the permitted lines (e.g.  the HI Balmer series).
The same species produces (at the same wavelengths) strong emission 
lines in the ionized gas and therefore the observed intensity of such lines is reduced by the underlying absorption, affecting the derived extinction.
%%%%%%%%%%%%%%%%%%%%%%%%%%%%%%%%%%%%%%%%%%%%%%%%%%%%%%%%%%%%%%%%%%%%%%%%%%%%%%%
For those galaxies for which the \hg\ line flux was avaliable, 
we estimate the correction due to the presence of
the low mass underlying population following Rosa-Gonz\'alez, Terlevich \& Terlevich~\cite{2002Rosa}.
The final values of Av (the extinction in the V wavelength given in magnitudes)
together with the \ha\ and \hb\ fluxes and the derived SFRs 
are presented in Table~\ref{tab:Opt}.

Notice that, as mentioned before, the SFR given by the \ha\ luminosity 
is tracing only the most massive stars (a few million years old), 
when a fraction of the expected HMXBs has yet to be formed.
The time delay between the formation of the stellar burst 
and the formation of the first HMXBs, 
is one of the main topics of this paper and  we will discuss it in detail in 
Section~\ref{sec:Ha}.

\begin{table*}
\caption{
Optical emission lines:  fluxes, visual extinctions and SFRs.}
\label{tab:Opt}
\begin{tabular}{lrcccr}
\hline
Name       & Distance  & F(H$\alpha$)$ (10^{-14})$ & F(H$\beta$) $(10^{-14})$ & Av     & SFR(H$\alpha$)  \\
\          & (Mpc)  &   (erg s$^{-1}$cm$^{-2}$)     &  (erg s$^{-1}$cm$^{-2}$) & \    & \msol yr$^{-1}$  \\  \hline
       Mrk52 &   31&    146.  $\pm$ 22.     & 17.0$\pm$2.5  &  3.06$\pm$0.46&    14.42$\pm$4.84  \\
Cam0902+1448 &  222&    28.1  $\pm$ 4.2     & 3.40$\pm$0.51 &  1.92$\pm$0.29&    50.48$\pm$10.92 \\
Tol1457-262  &   74&    85.8  $\pm$ 13.     & 19.4$\pm$2.9  &  1.50$\pm$0.22&    10.74$\pm$2.10  \\
 Tol1247-232 &  213&    50.4  $\pm$ 7.6     & 13.5$\pm$2.02 &  0.74$\pm$0.11&    36.52$\pm$12.25 \\
 Tol2259-398 &  127&    19.9  $\pm$ 3.0     & 3.09$\pm$0.46 &  2.25$\pm$0.34&    14.95$\pm$5.02  \\
   Cam08-82A &  222&     9.43 $\pm$ 1.4     & 1.10$\pm$0.16 &  3.05$\pm$0.46$\dag$ &    38.45$\pm$12.90 \\
      Mrk605 &  131&     8.07 $\pm$ 1.2     & 1.00$\pm$0.15 &  2.87$\pm$0.43$\dag$ &    10.15$\pm$3.40 \\
 Tol2306-400 &  292&     9.50 $\pm$ 1.4     & 0.87$\pm$0.13 &  3.71$\pm$0.56$\dag$ &   106.99$\pm$35.89 \\
      Mrk930 &   78&    39.4  $\pm$ 5.9     & 8.91$\pm$1.34 &  1.20$\pm$0.18&     5.34$\pm$1.79 \\
       UM530 &  316&    26.9  $\pm$ 4.0     & 4.37$\pm$0.65 &  2.12$\pm$0.32&   114.40$\pm$38.37 \\
 Tol0420-414 &   87&    86.0  $\pm$ 13.     & 28.2$\pm$4.23 &  0.17$\pm$0.02$\dag$ &     6.93$\pm$2.32 \\
 Tol0619-392 &  236&     9.53 $\pm$ 1.4     & 1.23$\pm$0.18 &  2.76$\pm$0.41&    35.72$\pm$11.98 \\
       UM421 &  177&     3.17 $\pm$ 0.48    & 0.24$\pm$0.04 &  4.24$\pm$0.64$\dag$ &    19.02$\pm$6.38 \\
       UM444 &  105&    17.1  $\pm$ 2.6     & 3.24$\pm$0.49 &  1.20$\pm$0.18&     4.15$\pm$1.39 \\
\hline 
\end{tabular}
{\footnotesize  
\\ Av  corrected by an underlying stellar population except when marked with a $\dag$ 
}
\end{table*}

\section{\xmm\ data analysis}

Fourteen objects have been  observed with  \xmm\  \cite{2001Jansen}.
The details  of the observations  are
summarized  in   Table~\ref{tab:xmmobs}.   The  raw   data  have  been
processed using the standard  {\it Science Analysis System},
SASv7.0.0~\cite{2004Gabriel}.  The most up-to-date  calibration files available
in July 2007 have been used  for the data reduction.  The raw data have
been filtered from high  background flaring using the method described
in Piconcelli et al.~\cite{2004Piconcelli}. Detailed spatial
and spectral analysis was possible only for four objects; net count
rates were derived for all targets in the sample (Table~\ref{tab:xmmobs}). None  of the
objects was  detected with the Reflection Grating Spectrometer.

\begin{table*}
\caption{\xmm\ observations  ID, total  and net exposure times after flaring removal,
and background subtracted count rates in the different X-ray bands.
The data were extracted from the {\it EPIC-pn} camera. }
\label{tab:xmmobs}
\begin{tabular}{llllll}
\hline
Target &  Obs. ID &  Exposure time  & Count Rate & Count Rate & Count Rate \\
 & &       &    (0.15 - 10 keV) &  (0.5 - 2 keV) &   (2 - 10 keV) \\
 & & ks & c/s & c/s & c/s  \\ \hline
Mrk 52            & 0303560101 & 6.84, 4.3 &  0.020$\pm$0.002	       &0.0132$\pm$0.0018	 &0.0008$\pm$0.0003    \\
Cam 0902+1448$\S$ & 3035620101 & 6.84, 3.3 &  0.110$\pm$0.011	       &0.053$\pm$0.006		 &0.030$\pm$0.008      \\
Tol 1457-262      & 0303560601 & 9.33, 6.5 &  0.059$\pm$0.004	       &0.040$\pm$0.003		 &0.013$\pm$0.003      \\
Tol 1247-232      & 0303561001 & 11.5, 6.0 &  0.024$\pm$0.003	       &0.0150$\pm$0.0019	 &0.0058$\pm$0.0014    \\
Tol 2259-398      & 0303560701 & 10.2, 5.6 &  0.010$\pm$0.003	       &0.0057$\pm$0.0019	 &0.004$\pm$0.003      \\
Cam 08-82A        & 0303561101 & 11.8, 6.7 &  $<$1.4$\times10^{-3}$    &$<5\times10^{-4}$	 &$< 6\times10^{-4}$   \\
Mrk 605           & 0303561701 & 7.93, 5.2 &  (1.9$\pm$1.5)10$^{-3}$   &$<9\times10^{-4}$	 &$< 9\times10^{-5}$   \\
Tol 2306-400      & 0303560401 & 10.9, 5.0 &  (8$\pm$3)10$^{-3}$       &(4.9$\pm$1.6)10$^{-3}$	 &0.003$\pm$0.002      \\
Mrk 930           & 0303560901 & 10.8, 2.4 &  (4.6$\pm$1.7)10$^{-3}$   &0.0017$\pm$0.0007	 &0.0029$\pm$0.0009    \\
UM 530            & 0303560501 & 6.84, 2.0 &  (8$\pm$3)10$^{-3}$       &(5$\pm$2)10$^{-3}$ 	 &$< 4\times10^{-4}$   \\
Tol 0420-414      & 0303561901 & 6.84, 2.8 &  (1.8$\pm$1.6)10$^{-3}$   &(1.4$\pm$0.9)10$^{-3}$	 &$< 3\times10^{-4}$   \\
Tol 0619-392      & 0303561401 & 6.84, 3.5 &  (5.2$\pm$2.3)10$^{-3}$   &3.0$\pm$1.4)10$^{-3}$	 &0.0020$\pm$0.0015    \\
UM 421            & 0303561601 & 11.8, 5.7 &  (2.9$\pm$1.7)10$^{-3}$   &0.0017$\pm$0.0011	 &0.0012$\pm$0.0009    \\
UM 444            & 0303561801 & 12.1, 3.6 &  (7$\pm$3)10$^{-3}$       &(4.8$\pm$1.5)10$^{-3}$   &$< 8\times10^{-4}$   \\
\hline
\end{tabular}\\
{\footnotesize 
$\S$ Another observation of Cam0902+1448 with ID 0303561201 is available.
The results on spectral quantities derived from it, though less accurate, 
are compatible with those obtained from the 3035620101 
observation so we used the latter for the analysis. }
\end{table*}

\subsection{Imaging Analysis}

\begin{figure}
\setlength{\unitlength}{1cm}           
\begin{picture}(7,8)       
\put(-1.,0.){\includegraphics{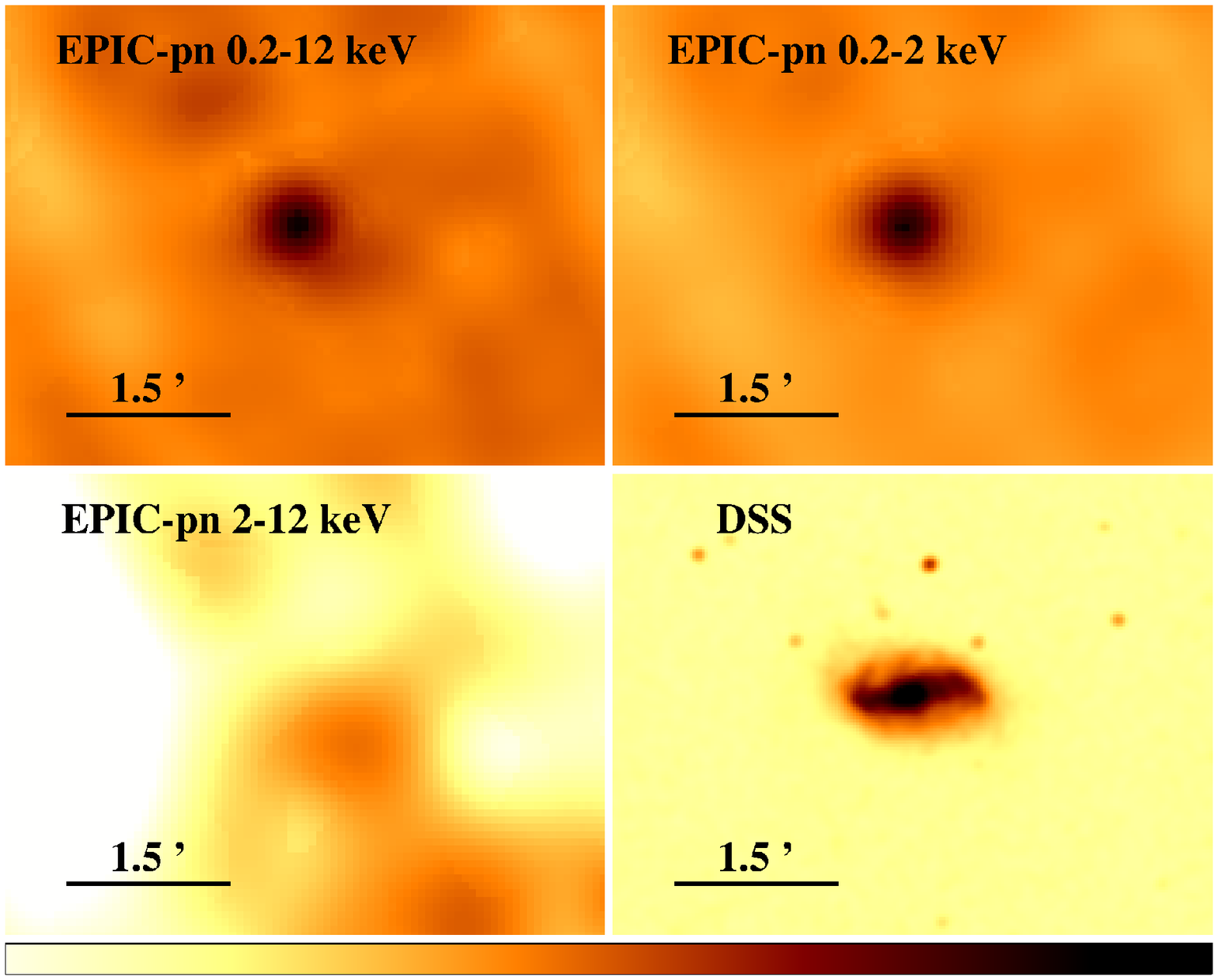}}
\end{picture}
\caption{\label{fig:mrk52} Smoothed  \pn\ images in the  0.2-12 keV, 0.2-2
keV, and 2-12~keV  energy bands of Mrk~52. We  also show the DSS
image for comparison. Here and in figures 2, 3, and 4, North is up and 
East is left.
}
\end{figure}

\begin{figure}
\setlength{\unitlength}{1cm}           
\begin{picture}(7,8)       
%%%% \put(-1.,-3.){\special{psfile=cam0902_image.ps
\put(-1.,-3.){\includegraphics{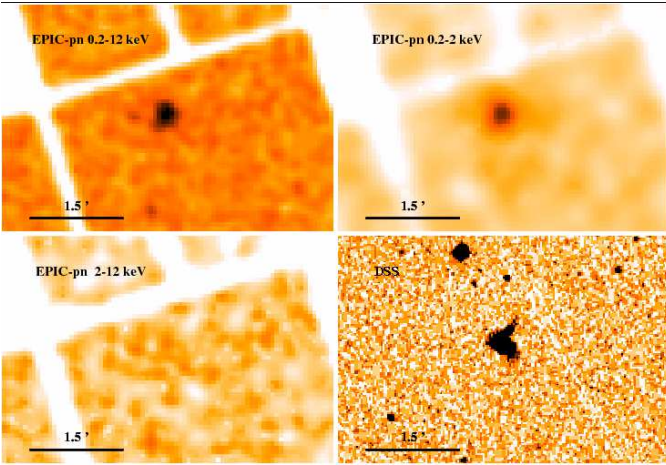}}
\end{picture}
\caption{\label{fig:cam0902} 
As in Figure 1, for  Cam~0902+1448.
}
\end{figure}

\begin{figure}
\setlength{\unitlength}{1cm}           
\begin{picture}(7,8)       
\put(-.75,0.){\includegraphics{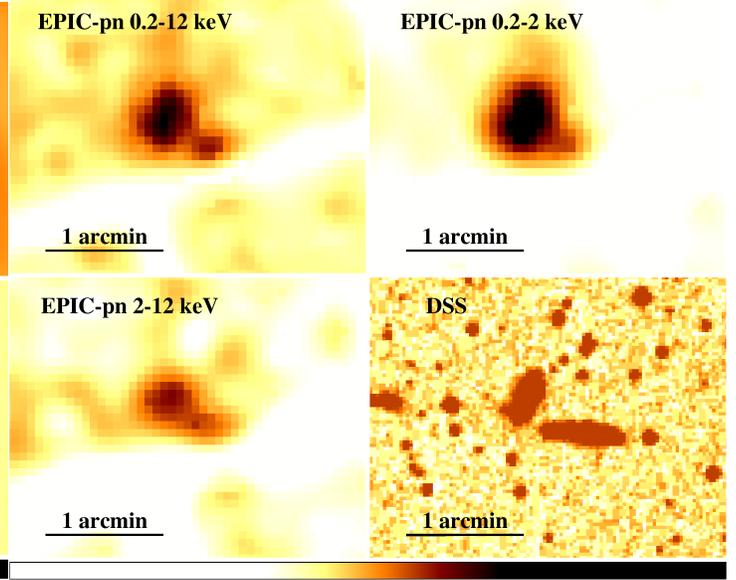}}
\end{picture}
\caption{\label{fig:tol1457} 
As in Figure 1, for  Tol~1457-262.
}
\end{figure}

\begin{figure}
\setlength{\unitlength}{1cm}           
\begin{picture}(7,8)       
\put(-0.75,0.){\includegraphics{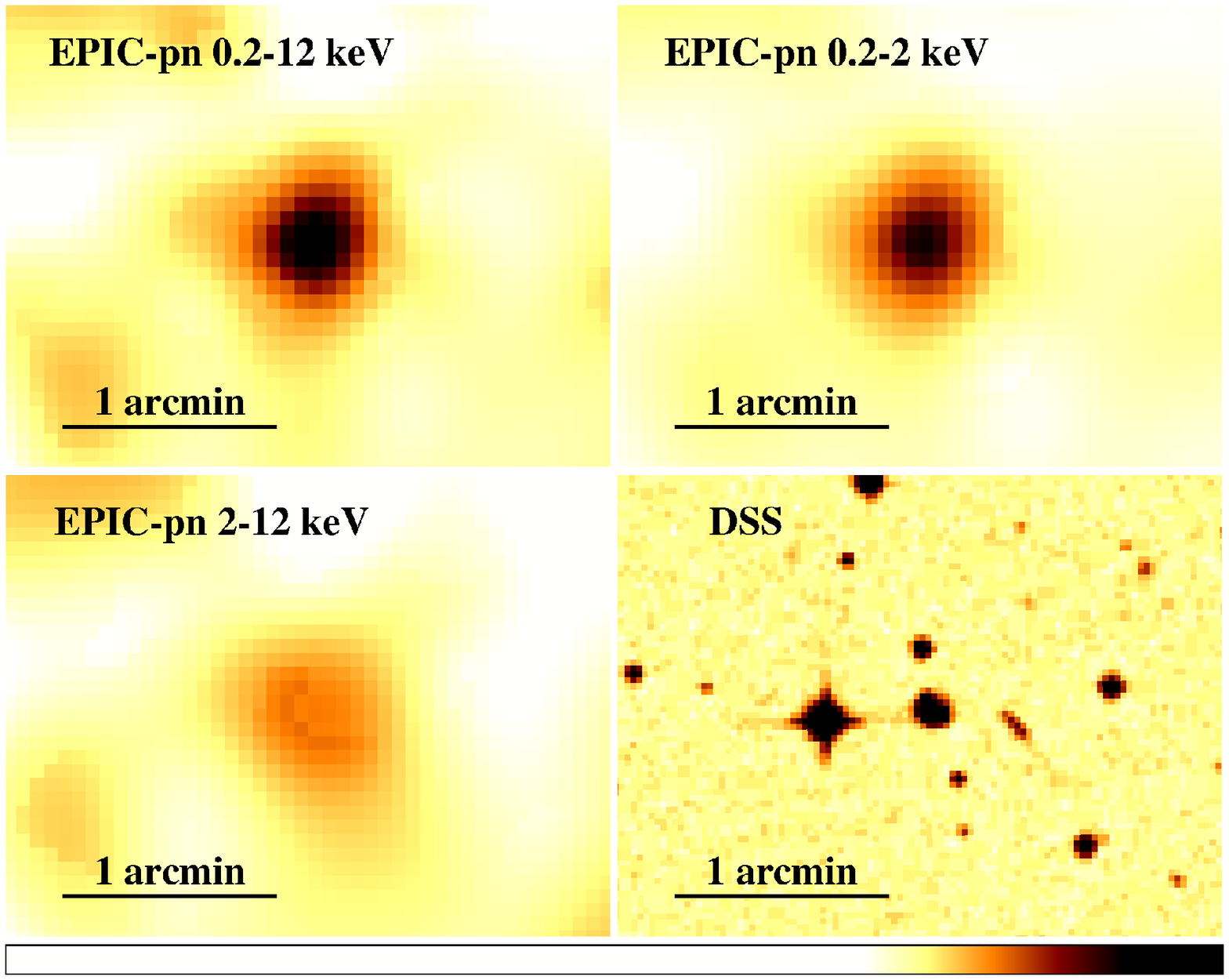}}
\end{picture}
\caption{\label{fig:tol1247} 
As in Figure 1, for  Tol~1247-232.
}
\end{figure}

The majority of the sources  are very weak in
the  X-ray  band, allowing  spatial  analysis  for  only four  of  the
galaxies:  Mrk~52, Tol~1457, Tol~1247, and Cam~0902.
We   have  generated   smoothed  images   for   these   galaxies
by    applying    the    SAS    task    
{\it asmooth}\footnote{The {\it  asmooth} task was applied  using the 
{\it adaptive} convolution technique with  a S/N=8.} to the \pn\ 
0.2-2~keV, 2-12~keV and 0.2-12~keV band images 
(Figures ~\ref{fig:mrk52},~\ref{fig:cam0902},~\ref{fig:tol1457}, and 
~\ref{fig:tol1247}).

The   images   of   Mrk~52  in   different   energy   bands
(Fig.~\ref{fig:mrk52})  show  a   soft  source,  with  no  significant
emission  in the  hard band.  This is  also the  case  of Tol~1247-232
(Fig.~\ref{fig:tol1247}), although  its emission  in the hard  band is
stronger that in Mrk~52. 
The images of Cam~0902+1448 in the different energy 
bands are shown in Fig.~\ref{fig:cam0902}. Tol~1457-262  is  actually  an  
HII galaxy pair. Both components  are detected in the X-ray  image
but could not be resolved (Fig~\ref{fig:tol1457}).  
The images show that the South-West member
is the weakest one of the pair, with a stronger contribution
in the  hard band,  i.e.  2-12~keV. 
%%The proximity of the two galaxies, and the morphology of the 
%%X-ray emission suggest an excess of emission due to 
%%the collision of the two giant HII regions.

%%%%%%%%%%%%%%%%%%%%%%%%% Figure 1 %%%%%%%%%%%%%%%%%%%%%%%%%%
%%%%%%%%%%%%%%%%%%%%%%%%%%%%%%%%%%%%%%%%%%%%%%%%%%%%%%%%%%%%%%%%%%%%

%%%%%%%%%%%%%%%%%%%%%%%%% Figure 2 %%%%%%%%%%%%%%%%%%%%%%%%%%
%%%%%%%%%%%%%%%%%%%%%%%%%%%%%%%%%%%%%%%%%%%%%%%%%%%%%%%%%%%%%%%%%%%%

%%%%%%%%%%%%%%%%%%%%%%%%% Figure 3 %%%%%%%%%%%%%%%%%%%%%%%%%%
%%%%%%%%%%%%%%%%%%%%%%%%%%%%%%%%%%%%%%%%%%%%%%%%%%%%%%%%%%%%%%%%%%%%

%%%%%%%%%%%%%%%%%%%%%%%%% Figure 4 %%%%%%%%%%%%%%%%%%%%%%%%%%
%%%%%%%%%%%%%%%%%%%%%%%%%%%%%%%%%%%%%%%%%%%%%%%%%%%%%%%%%%%%%%%%%%%%

\subsection{Spectral analysis}
\label{SpecAnalysis}

For the majority of the sources (ten out of fourteen), the limited
signal--to--noise did not allow spectral analysis with the EPIC camera. 
However, enough counts were detected to produce spectra for 
four sources: Mrk~52, Cam~0902, Tol~1457, and Tol~1247.
Circular extraction regions of 400, 700, 550 and 650 pixels respectively
were selected in order to maximize signal--to--noise in  the 0.2-10~keV band 
(see details in Piconcelli et al.~\cite{2004Piconcelli}.
%%%%%%%%%%%%%%   New text %%%%%%%%%%%%%%%%%%%%%%%%%%%%%%%%%%%%%%%%%%
By using these regions we are including between 80\% and 90\% of
the total energy as described in the \xmm\, handbook. We also checked that
we are not including X-ray emission from close sources and
from pixels too close to the CCD edges.

%%%%%%%%%%%%%%%%%%%%%%%%%%%%%%%%%%%%%%%%%%%%%%%%%%%%%%%%%%%%%%%%%%%%
Background spectra  have  been  extracted from a circular region located 
in the frame of the galaxy and free of any visible contaminating
source. 
The  associated ancillary and response  matrices were obtained
using the standard SAS tasks.   {\it EPIC-}\pn\ and the combined \mos\
1 and  2 spectra  of the four  brightest objects were  extracted.  The spectra of
Cam~0902+1448, Tol~1457-262, and Tol~1247-232   have been binned
such that each  bin contains at least 20 counts in  order to apply the
$\chi^2$ minimization  technique.  As not enough  counts were detected
in  Mrk~52,  we used  the  unbinned  spectra to  perform  the
analysis  and  therefore   C-statistics  were  applied.  The \xmm\ spatial
resolution  and  the limited  signal--to--noise  do  not  allow us  to
spectroscopically  analyze each component of  the  pair Tol~1457
separately.  Both  members are included  in the extraction  region and
therefore fluxes and luminosities  refer to the pair's combined emission.
In all the galaxies, the  spectral
analysis has  been performed using  the v12.0 of {\small  XSPEC}.  The
errors  quoted  are  referred  to  the  90\%  confidence  level,  i.e.
$\Delta\chi=2.71$ when $\chi^2$ statistics were applied.
In all the cases, we fixed the Galactic column density to the values listed in
Table~\ref{tab:FluxesAndLum} which were obtained through the 
Leiden/Argentine/Bonn Survey of Galactic HI~\cite{2005Kalberla}.

The high  supernova rate occurring in star-forming  regions produces an
acceleration of  the electrons to relativistic  velocities. The
scattering of  these electrons with  FIR photons, enhanced  by the
star-forming  process results  in  non-thermal X-ray  emission.  This
emission can, at first order, be  modelled as a power law with an index
in the range 1.6-1.8.  However, observational  studies on star-forming
galaxies find that  a thermal emission model with  temperatures of the
order of 8~keV,  provides also a satisfactory representation of the observed
high energy spectra. The average X-ray spectrum of a population of
low- and high-mass X-ray binaries  can be described by a power law with an index
$\Gamma\sim1.2$ and a cut-off around 7.5~keV~\cite{2003Persic}. 

On the other hand, the  diffuse emission mainly  contributes to
the soft X-ray band and can also be modelled as a thermal component but
with a temperature one order  of magnitude lower.  Bearing this 
picture in mind, we  have tested  three different  models to  fit  the source
spectra: a single  power law, a thermal emission  model [specifically
the  {\it  mekal}  model in  {\small  XSPEC},
Liedahl, Osterheld \& Goldstein~\cite{1995Liedahl}],  and  a
combination of  the best  of the two previous  ones with a  second thermal
emission model accounting for the soft emission. 
We also included a model for the photo electric  absorption ({\it  zwabs} in
{\small  XSPEC}) using the Wisconsin cross-sections~\cite{1983Morrison}.

\begin{figure*}
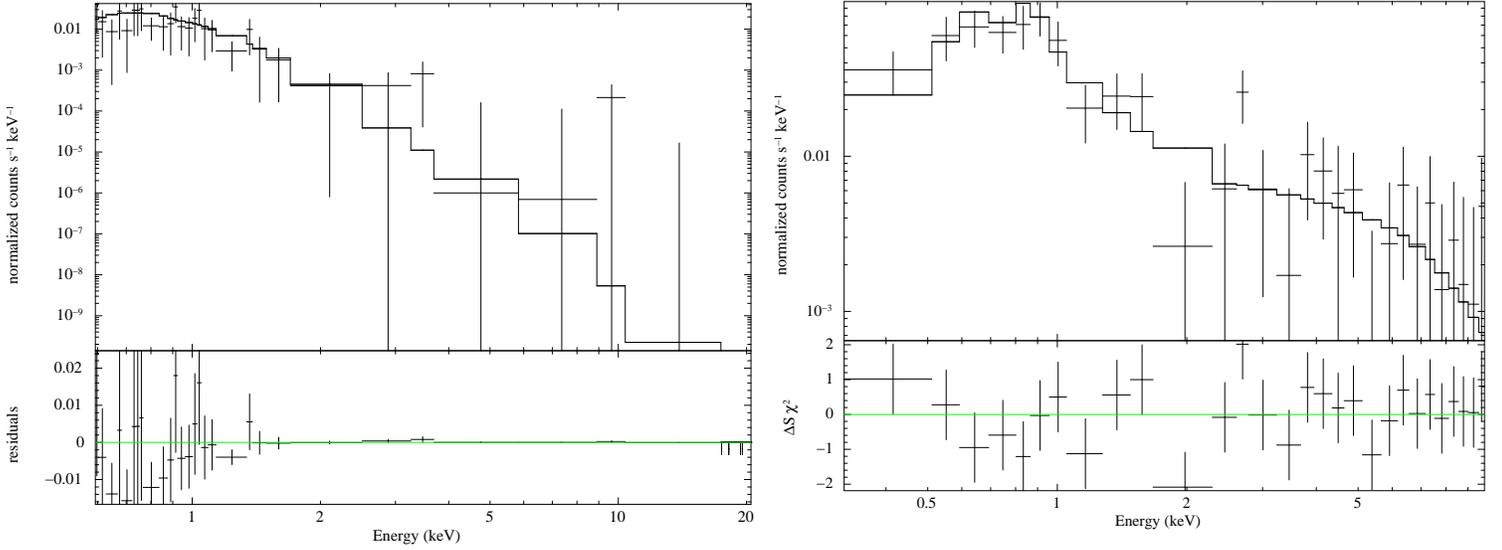

\setlength{\unitlength}{1cm}           
\begin{picture}(7,8)       
\put(-6.5,8.5){\includegraphics{mrk52_updated.ps}}
\put(3.5,8.5){\includegraphics{cam0902_spec.ps}}
\end{picture}
\caption{\label{fig:spec_cam0902_mrk52} 
Observed {\it EPIC} spectra, best fit model and residuals of {\it left panel:} 
Mrk~52;  {\it right panel:} Cam~0902+1448. 
}
\end{figure*}

\begin{figure*}
\setlength{\unitlength}{1cm}           
\begin{picture}(7,7.8)       
\put(-7.,8.2){\includegraphics{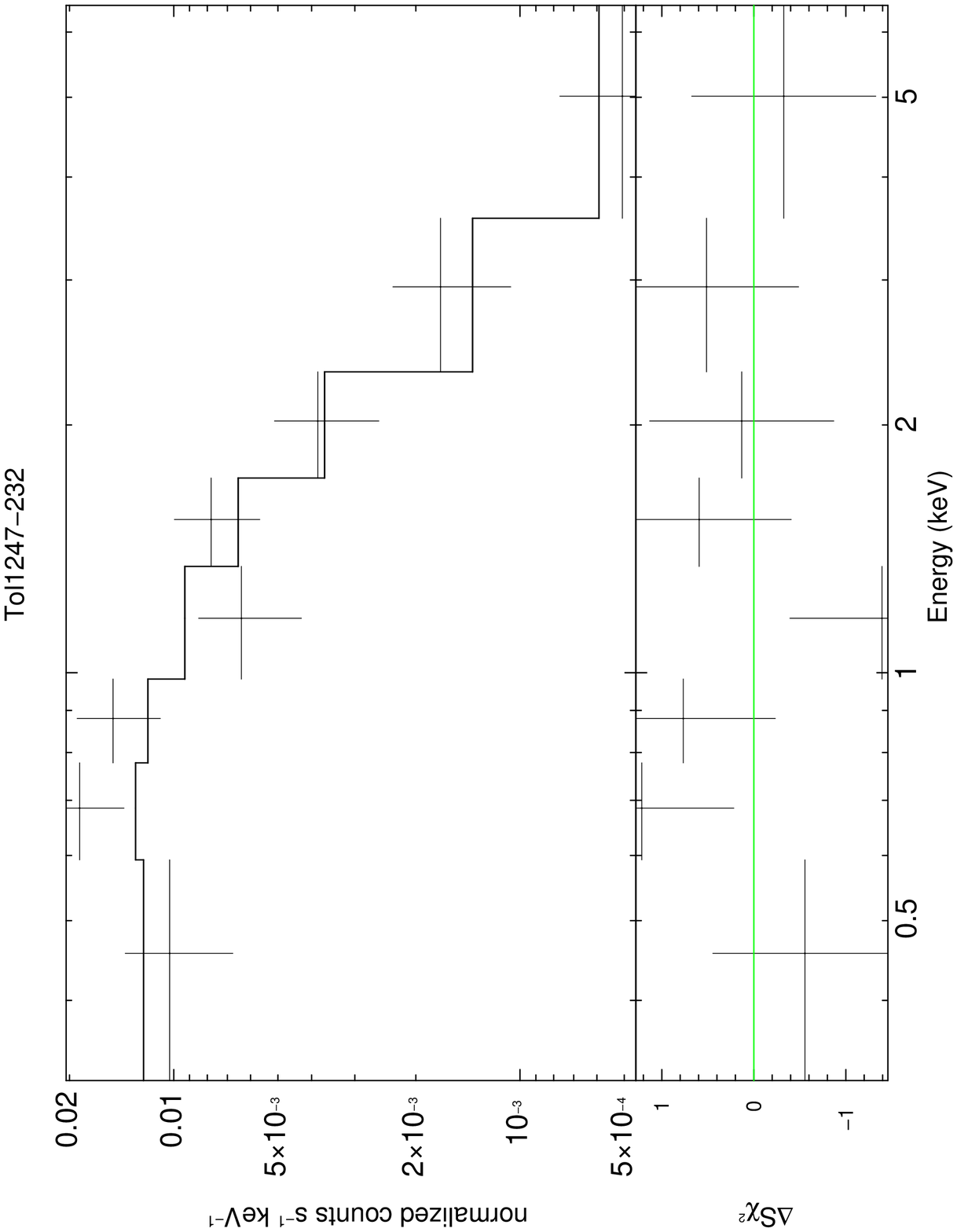}}
\put(3.25,8.2){\includegraphics{tol1457_spec.ps}}
\end{picture}
\caption{\label{fig:spec_tol1247_tol1457}  As in Figure 5 for
{\it left panel:} Tol1247-232; 
{\it right panel:} Tol1457-262. 
}
\end{figure*}

\begin{figure}
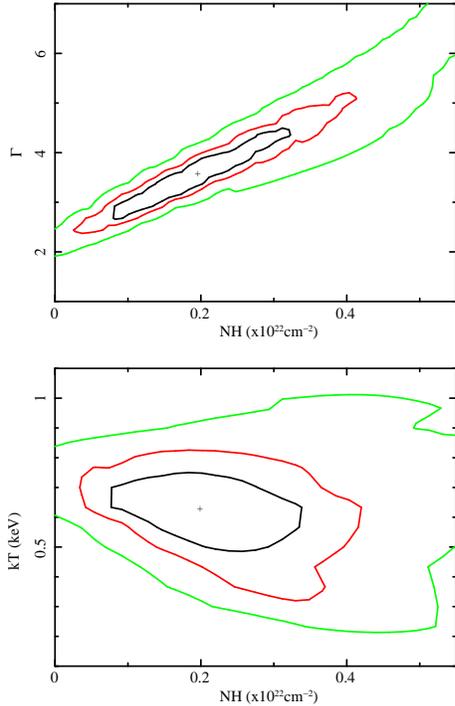

\setlength{\unitlength}{1cm}           
\begin{picture}(6,9)       
\put(1.0,9.75){\includegraphics{cont_cam09_gamma.ps}}
\put(1.0,4.9){\includegraphics{cont_cam09_kt.ps}}
\end{picture}
\caption{\label{fig:contour} 
Confidence contours for the determination of 
$\Gamma$ and $N_H$ (top panel) and temperature and $N_H$  (bottom panel) for 
Cam0902. In both panels the best fit is marked with a cross. 
}
\end{figure}

The parameters fitted -- by using a Levenberg-Marquart method -- 
and the  goodness of the  fits are given in  Table~\ref{tab:fits}. 
The low  number of  detected  counts does  not  allow us  to fit more  complex models.   
A single-component model provides acceptable fits to the data for all sources
analyzed, except for Cam~0902, for which a two components fit is best.  
The  derived fluxes  and  unabsorbed  luminosities for each
object are given in Table~\ref{tab:lum}.
In Figs~\ref{fig:spec_cam0902_mrk52}  and~\ref{fig:spec_tol1247_tol1457},
we illustrate  the observed  spectra and the best fit models for each
galaxy.   
%%%%%%%%%%%%%% New text %%%%%%%%%%%%%%%%%%%%%%%%%%%%%%%%%%%%%%%%%%%%%%%%%%%
To illustrate the variation of the errors in the
$\Gamma$ vs. $N_H$ and the temperature vs. $N_H$ planes
we plotted in Figure~\ref{fig:contour} the corresponding error
ellipses for Cam~0902.
%%%%%%%%%%%%%%%%%%%%%%%%%%%%%%%%%%%%%%%%%%%%%%%%%%%%%%%%%%%%%%%%%%%%%%%%%%%%

%%%%%%%%%%%%%%%%%%%%%%%%% Figure 5 %%%%%%%%%%%%%%%%%%%%%%%%%%
%%%%%%%%%%%%%%%%%%%%%%%%%%%%%%%%%%%%%%%%%%%%%%%%%%%%%%%%%%%%%%%%%%%%

%%%%%%%%%%%%%%%%%%%%%%%%% Figure 6 %%%%%%%%%%%%%%%%%%%%%%%%%%
%%%%%%%%%%%%%%%%%%%%%%%%%%%%%%%%%%%%%%%%%%%%%%%%%%%%%%%%%%%%%%%%%%%%

\begin{table*}
%\begin{sidewaystable}[hb]

\caption{Spectral  analysis.}\label{tab:fits}
\begin{tabular}{lllllll}
\hline 

 {\bf Target} & {\bf Model} & N$_H$ & $\Gamma$ &    N$_H$ &kT  & Goodness \\
& (1) \\
& & 10$^{21}$cm$^{-2}$ & &  10$^{21}$cm$^{-2}$ & keV \\

\hline\hline
Mrk 52 & {\bf A} & $6\pm3$ &  $6.5\pm0.5$ & - & - & {\bf 489 for 3222 dof$^\ddagger$}\\
 & A$\dagger$ & 5.5f & $6.5\pm0.5$ & - & - & 490 for 3223 dof$^\ddagger$\\
 & B & - & - & $7.7^{+1.5}_{-1.3}$ & $0.14\pm0.02$ & 497 for 3222 dof$^\ddagger$\\
 & B$\dagger$ & - & - &  5.5f & $0.18^{+0.02}_{-0.03}$ &  505 for 3223 dof$^\ddagger$\\
\hline
%Cam 0902+1448 & A & $3.4^{+2.0}_{-1.6}$ & $4.4^{+1.4}_{-1.0}$ & - & - & 40 for 35 dof\\
%& A$\dagger$ & 5.2f & $5.3^{+0.3}_{-0.4}$ & - & - & 42 for 36 dof\\
%& B & - & -  & $8.4\pm1.1$ & 0.18$^{+0.02}_{-0.03}$ & 62 for 35 dof \\
%& B$\dagger$ & - & - & 5.2f & 0.25 & 79 for 36 dof \\
%& {\bf C} & $7.7^{+1.0}_{-0.9}$ & 1.0$^{+0.8}_{-0.5}$ & - & 0.147$^{+0.03}_{-0.011}$ & {\bf 22 for 33 dof}\\ 
\hline 
Cam 0902+1448          & A & 4.6$^{+1.8}_{-1.2}$ & 5.2$^{+0.8}_{-1.2}$ & -   & -                   & 31 for 25 dof\\
                       & A$\dagger$ & 3.4f & $5.3^{+0.3}_{-0.4}$ & - & - & 35 for 26 dof\\
                       & B & -                   & -                   & 0.8 & 0.2                 & 63 for 25 dof\\
                       & B$\dagger$ & - & - & 3.4f & 0.25 & 73 for 26 dof \\
                       & {\bf C} & 2.0$^{+1.5}_{-1.6}$  & 3.6$^{+1.2}_{-1.1}$ & -   & 0.63$^{+0.16}_{-0.17}$ &  {\bf 18 for 23 dof}\\
\hline
Tol1457-262 & {\bf A } & $1.4^{+1.0}_{-0.9}$ & $2.2^{+0.5}_{-0.4}$ & - & - & {\bf 16.2 for 12 dof} \\
& B & - & - & $<0.6$ & $4.6^{+3}_{-1.5}$ & 18.7 for 12 dof\\
% & C & $7.1^{+2}_{-3}$ & $2.6\pm0.4$ & - & $0.11^{+0.04}_{-0.03}$ & 35 for 32 dof \\
\hline

Tol1247-232 & {\bf A} & $<1.4$& $1.8^{+0.4}_{-0.3}$ & - & - & {\bf 5.1 for 6 dof} \\
& B & - & - & $<0.5$ & $7^{+16}_{-4}$ & 7.6 for 6 dof \\
%& C & - & $1.5\pm0.3$ & - & $>0.1$ & 2.4 for 4 dof \\
\hline
\end{tabular}

{\scriptsize Notes: (1) Model A: zwabs*zpowerlaw;Model B: zwabs*mekal; Model C: zwabs(zpowerlaw+mekal);  in boldface, the best fit model.$\dagger$ N$_{\rm H}$ fixed to the value
derived from A$_{\rm V}$.  $^\ddagger$ C-statistics. 
  }
\end{table*}

\begin{table*}
\caption{Observed fluxes and unabsorbed  luminosities in the soft and hard
bands.}\label{tab:lum}
\begin{tabular}{lllll}
\hline
{\bf Target} & {\bf Flux (0.5-2 keV)} & {\bf Flux (2-10 keV)} & {\bf Luminosity (0.5-2 keV)} & {\bf Luminosity (2-10 keV)} \\
 & 10$^{-14}$ erg cm $^{-2}$s$^{-1}$ & 10$^{-14}$ erg cm $^{-2}$s$^{-1}$ & 10$^{41}$ erg s$^{-1}$ & 10$^{41}$ erg s$^{-1}$ \\
\hline\hline

Mrk 52 & $2.5\pm0.5$ & 0.08$\pm$0.03 & $5.8\pm1.2$ & $1.0\pm0.4\times10^{-3}$\\
Cam0902+1448& $8.5^{+1.5}_{-4}$ & 1.5$^{+0.4}_{-1.3}$& 14$^{+2}_{-7}$  & 1.1$^{+0.3}_{-0.9}$\\
Tol1457-262 & $5.5^{+1.3}_{-1.9}$ & $8.5^{+1.5}_{-2}$ & $0.7\pm0.2$ & $0.58^{+0.11}_{-0.14}$ \\
Tol1247-232 & $2.2^{+0.8}_{-1.3}$ & $4.4^{+0.7}_{-2}$ & $1.6^{+0.6}_{-0.9}$ & $2.5^{+0.4}_{-1.1}$\\
\hline
\end{tabular}
\end{table*}

We compare the obtained X-ray fluxes (Table~\ref{tab:lum}) with those 
given by a starburst model described below (Table~\ref{tab:FluxesAndLum}). 
The agreement is good but some discrepancies were
found in the soft X-ray fluxes mainly due to the uncertainties in the N$_H$ value.
In fact, differences in the goodness of the fit were found when
the fit was made by fixing the value of the  N$_H$ derived from the Hydrogen
emission lines ($N_H=1.79\times 10^{21}~\times$~Av, cm$^{-2}$) as compared
with the results obtained when the N$_H$ is a free parameter of the
fitting procedure.
The derived values of $\Gamma$ are somewhat higher than
those found in the literature for similar type of objects and 
fix $\Gamma= 2$ created fits with large values of $\chi^2$.
However the data quality does not allow the discussion of the 
origin of this apparent systematic effect.
In one case (Cam~0902), the best fit was obtained by adding 
to the power law component a thermal model with a temperature of about 0.6 keV.
That is consistent with temperatures of the hot plasma observed in other star forming galaxies
\cite{2005Grimes}.

%%%%%%%%%%%%%%%%%%%%%%%%%%New text %%%%%%%%%%%%%%%%%%%%%%%%%%%%%%%%%%%%%%%%%%%%%%
In order to have a homogeneous data set, and due to the low number of 
photons that could be affecting the conclusions obtained from the spectra, 
the analysis presented in the next sections is based only on the count rates 
from Table~\ref{tab:xmmobs}. 
%%%%%%%%%%%%%%%%%%%%%%%%%%%%%%%%%%%%%%%%%%%%%%%%%%%%%%%%%%%%%%%%%%%%%%%%%%%%%%%%%
The count rates were calculated for  the \pn\ detector, extracted  from a
region of 15\arcsec\ free  of contaminating sources and corrected for
background.  
We convert the count rates by using a two component model;
a power law with spectral index of 1.2  and a cut-off at 7.5 keV (associated to
the  combined high  and low-mass  X-ray binaries  contribution)  plus a
thermal  emission  model with  kT=0.8~keV  (associated  to the  diffuse
emission from galactic winds). 
%%%%%%%%%%%%%%%%%%%%%%%%%%%% New text %%%%%%%%%%%%%%%%%%%%%%%%%%%%%%%%%%%%%%%%%%
The two components (power law and thermal emission) are normalized  
such that the flux ratio between the thermal and the 
power law  components in the 2---10 keV band is 0.03
[\cite{2003Persic}, and references therein].   
%%%%%%%%%%%%%%%%%%%%%%%%%%%%%
The estimated  fluxes  in the  soft  (0.5-2~keV) and  hard
(2-10~keV) bands  are shown in  Table~\ref{tab:FluxesAndLum}.  
Calculations were  performed using {\it  PIMMS}v3.9\footnote{PIMMS does
not   consider  the   cut-off  power-law  for   fluxes/count  rates
estimates.  Therefore, we have also applied a correction
due to differences on flux  estimation when a simple power-law model is used
instead of a cut-off power law.}. 

%%%%%%%%%%%%%%%%%%%%%%%%%%%% Changed text %%%%%%%%%%%%%%%%%%%%%%%%%%%%%%%%%%%%%%%%%%
The fluxes and luminosities, also given in Table~\ref{tab:FluxesAndLum}, were corrected
from  absorption using  the extinction  derived from published ratios between 
\ha\ and \hb\, which include the absorption due to the Galaxy. Notice that 
the values of the column density given in Table~\ref{tab:FluxesAndLum} 
were used only in the spectral analysis of Mrk52, Cam0902, Tol1457 and Tol1247.
%%%%%%%%%%%%%%%%%%%%%%%%%%%%%%%%%%%%%%%%%%%%%%%%%%%%%%%%%%%%%%%%%%%%
The errors were derived from the count rate uncertainties.  However, 
we notice that they should be considered lower limits of the real
uncertainties, as they only include the poisson  error on the count rate.
The uncertainty generated by considering different models is not included in the
error budget. In this sense, when a single power-law model 
with  $\Gamma=2$ is applied  instead of  the cut-off power-law plus
thermal emission, as  suggested  by  other  studies on  single  sources [e.g.
NGC~3310, NGC~3690 by Zezas, Georgantopoulos, \& Ward~\cite{1998Zezas}; 
NGC~253, M82 by Cappi et al.~\cite{1999Cappi}],
differences  of the  order of  20 percent in  the estimation  of  fluxes are
obtained. 
%%%%%%%%%%%%%%%%%%%%%%%%%%%% New text %%%%%%%%%%%%%%%%%%%%%%%%%%%%%%%%%%%%%%%%%%
The metallicity of the gas is another source of uncertainty
particularly in the
soft band. Calculations realized with {\it  PIMMS} show that
for a plasma with a temperature of 0.8 keV a change in metallicity from
solar to 20\% solar, typical of HII galaxies
[e.g. Hoyos~\& D{\'{\i}}az ~\cite{2006Hoyos}], produces about 15\% variation in the fluxes.
%%%%%%%%%%%%%%%%%%%%%%%%%%%%%%%%%%%%%%%%%% %%%%%%%%%%%%%%%%%%%%%%%%%%%%%%%%%%%%%
%%%%%%%%%%%%%%%%%%%%%%%%%%% New text %%%%%%%%%%%%%%%%%%%%%%%%%%%%%%%%%%%%%%%%%%%%
Finally, we estimate the luminosity due to low mass X-ray binaries
(L$_{LMXB}$) that could
be present in the selected galaxies. Based on 2MASS K-band magnitudes
we confirm that the observed galaxies have masses which are in general
lower than
that of an L$_\star$ galaxy (assuming a mass to light ratio in the
K-band of 0.5,
Bell \& de Jong~\cite{2001Bell}.
The contribution of these binaries to the observed X-ray luminosities 
is estimated using the relation between the stellar mass and the
luminosity due to LMXB
(Grimm, Gilfanov \& Sunyaev \cite{2002Grimm},
\begin{equation}
L_{LMXB}\,(erg\, s^{-1}) =  5 \times 10^{28} M_\ast
\end{equation}
where $M_\ast$ is the stellar mass in solar units. 
Table~\ref{tab:FluxesAndLum} shows that the calculated $L_{LMXB}$ are 
much smaller, on average less than 2 percent and at most 20 percent, than 
the observed X-ray luminosities.
The general result is that the total X-ray luminosities of the low mass galaxies selected here 
is probably not greatly affected by the LMXBs contribution.

%%%%%%%%%%%%%%%%%%%%%%%%%%%%%%%%%%%%%%%%%%%%%%%%%%%%%%%%%%%%%%%%%%%%%%%%%%%%%%%%%

\begin{table*}
\caption{Galactic column density and derived fluxes and luminosities.
}
\label{tab:FluxesAndLum}
\begin{tabular}{lclllllr}
\hline
Target & N$_H$  & 
F$_{0.5-2\,keV}$  & F$_{2-10\,keV}$& L$_{0.5-2\,keV}$&  L$_{2-10\,keV}$ &  L$_{0.5-2\,keV}^{thermal}$ & L$_{LMXB}$\\ 
 & ($10^{20}$)    &   $(10^{-14})$ &  $(10^{-14})$ & $(10^{41})$ & $(10^{41})$ & $(10^{41})$ & $(10^{39})$\\
 & cm$^{-2}$& erg/s/cm& erg/s/cm  & erg/s& erg/s & erg/s & erg/s \\ \hline
Mrk 52            &1.9  &    8.40$\pm$0.88&  9.90$\pm$1.04&  0.10$\pm$ 0.01 &      0.11$\pm$ 0.01 &  0.048$\pm$0.005&    1.76$\pm$0.53\\
Cam 0902+1448$\S$ &3.7  &   41.00$\pm$4.13& 48.00$\pm$4.84& 24.24$\pm$ 2.44 &     28.38$\pm$ 2.86 & 12.1  $\pm$1.22 &    2.34$\pm$0.70\\
Tol 1457-262      &9.4  &   20.00$\pm$1.36& 24.00$\pm$1.63&  1.30$\pm$ 0.09 &      1.56$\pm$ 0.11 &  0.651$\pm$0.044&    0.47$\pm$0.14\\
Tol 1247-232      &6.6  &    6.40$\pm$0.77&  7.50$\pm$0.90&  3.48$\pm$ 0.42 &      4.07$\pm$ 0.49 &  1.73$\pm$0.209&$<$ 0.73         \\
Tol 2259-398      &1.1  &    3.70$\pm$1.11&  4.30$\pm$1.29&  0.71$\pm$ 0.21 &      0.83$\pm$ 0.25 &  0.357$\pm$0.107&    1.04$\pm$0.31\\
Cam 08-82A        &3.3  &    $<$0.43      &  $<$0.50      &  $<$0.25        &$<$      0.30        &$<$ 0.127        &$<$ 0.63         \\
Mrk 605           &1.9  &    $<$0.82      &  $<$0.97      &  $<$0.17        &$<$      0.20        &$<$ 0.085        &    1.36$\pm$0.41\\
Tol 2306-400      &1.3  &    3.73$\pm$1.40&  4.30$\pm$1.61&  3.81$\pm$ 1.43 &      4.39$\pm$ 1.65 &  1.90 $\pm$0.714&    3.96$\pm$1.19\\
Mrk 930           &5.5  &    1.40$\pm$0.52&  1.61$\pm$0.60&  0.10$\pm$ 0.04 &      0.12$\pm$ 0.04 &  0.051$\pm$0.019&$<$ 0.08         \\
UM 530            &1.5  &    2.74$\pm$1.03&  3.18$\pm$1.19&  3.27$\pm$1.22  &$<$      3.79        &  1.63 $\pm$0.612&    6.43$\pm$1.93\\
Tol 0420-414      &2.5  &    0.58$\pm$0.52&  0.69$\pm$0.61&  0.05$\pm$0.05  &$<$      0.06        &  0.026$\pm$0.023&$<$ 0.13         \\
Tol 0619-392      &7.6  &    2.20$\pm$0.97&  2.60$\pm$1.15&  1.47$\pm$ 0.65 &      1.73$\pm$ 0.77 &  0.734$\pm$0.325&    1.78$\pm$0.53\\
UM 421            &4.1  &    1.50$\pm$0.88&  1.80$\pm$1.06&  0.56$\pm$ 0.33 &      0.67$\pm$ 0.39 &  0.280$\pm$0.164&    2.28$\pm$0.69\\
UM 444            &2.1  &    2.20$\pm$0.94&  2.60$\pm$1.11&  0.29$\pm$0.12  &$<$      0.34        &  0.144$\pm$0.062&    0.54$\pm$0.16\\
\hline
\end{tabular}
\end{table*}

\section{Early Evolutionary Phases: Comparison with SFR tracers}  
\label{sec:Ha}
\subsection{\ha\ luminosities}

\begin{figure}
\setlength{\unitlength}{1cm}           
\begin{picture}(7,6.75)       
\put(-1.5,-7.25){\includegraphics{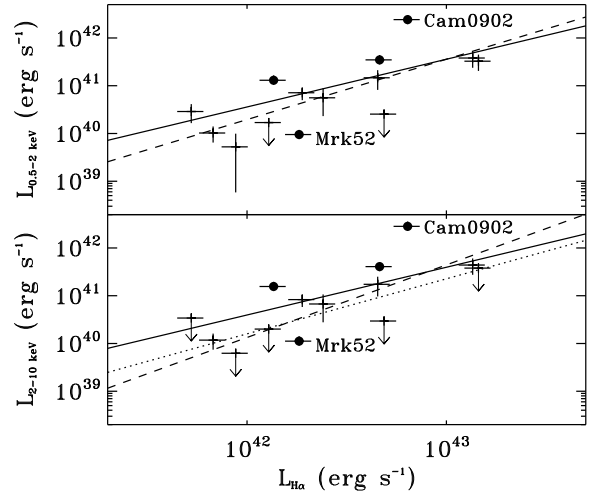}}
\end{picture}
\caption{\label{fig:lx_lalpha} 
X-ray (soft band in the top and hard band in the bottom panel)
vs.~\ha\ luminosities. Here and in the following plots,
solid circles mark 
those galaxies for which we extracted the X-ray spectrum. 
In both cases, the solid line is the result of converting the 
Ranalli law to  \ha\ luminosities and the  dashed one is a
linear fitting that takes into account the presence of upper limits in the 
derived X-ray luminosities. In the bottom panel, the dotted line 
is the result of  the fit obtained when leaving out the outlier Mrk52.
}
\end{figure}

The recombination emission lines only appear during the early 
evolution of a starburst when the massive stars are hot 
enough to produce ionizing photons. 
Therefore, the deduced SFR(\ha) is tracing  
the first million years of the present starburst evolution. 
Figure~\ref{fig:lx_lalpha}\ shows the relation between 
the \ha\ luminosity and the X-ray luminosities
both in the soft (top panel) and in the hard (bottom panel) bands.
The solid lines show the empirical relation
found by Ranalli et al.~\cite{2003Ranalli}
based on the study of nearby star forming galaxies, 

\begin{equation}
SFR_{soft}(M_\odot yr^{-1}) = 2.2 \times 10^{-40} L(0.5-2\, keV)\, (erg\, s^{-1}) 
\end{equation}
\begin{equation}
SFR_{hard}(M_\odot yr^{-1}) = 2. \times 10^{-40} L(2-10\, keV)\, (erg\, s^{-1})
\end{equation}

There is a good correlation between the L(\ha ) and the soft X-ray luminosity 
(top panel), but there is one galaxy (Cam~0902+1448) for which 
the luminosity in X-rays is well above  the one predicted by Ranalli's law. 
In the bottom plot of this Figure [hard X-rays vs.~L(\ha )]
most of the galaxies are below the empirical correlation.
%%%%%%%%%%%%%%%%%%%%%%%%%%%  New Text %%%%%%%%%%%%%%%%%%%%%%%%%%%%%%%%%%%%%%%%%%
Due to the presence of upper limits we fit the luminosities by using  
survival analysis\footnote{See IRAF task stsdas.analysis.statistics.survival for an introduction
to survival analysis.}. 
The Buckley-James algorithm [BJ, e.g.~Isobe, Feigelson, \& Nelson~\cite{1986Isobe}]
a non-parametric method which treats the residuals 
using the Kaplan-Meier description,  was applied to the data
in Figure~\ref{fig:lx_lalpha}.  
In the case of the bottom panel, the non-parametric method was applied both to all the
data (dashed---line) and excluding Mrk52 (dotted---line).
%%%%%%%%%%%%%%%%%%%%%%%%%%%%%%%%%%%%%%%%%%%%%%%%%%%%%%%%%%%%%%%%%%%%%%%%%%%%%%%%

%%%%%%%%%%%%%%%%%%%%%%%%% Figure 7 %%%%%%%%%%%%%%%%%%%%%%%%%%
%%%%%%%%%%%%%%%%%%%%%%%%%%%%%%%%%%%%%%%%%%%%%%%%%%%%%%%%%%%%%%%%%%%%

\subsection{Radio luminosity}

Table~\ref{tab:RadioInfraUV} shows the radio fluxes at 1.4 GHz for the
selected galaxies extracted from NED. The strong correlation known to exist between the radio and 
the far infrared fluxes is taken as an indication that radio emission is a 
reliable tracer of the SFR. Yun, Reddy, 
\& Condon~\cite{2001Yun} found, from a FIR selected sample of about 1800 star-forming 
galaxies, that their radio continuum may be used to  infer the extinction-free
SFR using the relation

\begin{equation}
SFR(1.4\,GHz) (M_\odot yr^{-1}) = 5.9\times 10^{-29}  L(1.4\,GHz) (erg\,s^{-1}Hz^{-1})
\end{equation}
The  values of SFR(1.4\,GHz) for our \xmm\ observed galaxies are also 
presented in Table~\ref{tab:RadioInfraUV}.
The empirical relation found by Yun et al.~\cite{2001Yun}
is well understood and it is based on the existence of a star forming 
event in which the observed radio emission arises
from accelerated electrons produced in the supernova remnants 
and the IR radiation comes from dust that is heated mainly by massive stars. 
The estimated SFR corresponds to continuous star formation occurring
for the last 10$^8$ years.  
%%%%%%%%%%%%%%%%%% Modified text %%%%%%%%%%%%%%%%%%%%%%%%%%%%%%%%%%

%%%%%%%%%%%%%%%%%%%%%%%%% Figure 8 %%%%%%%%%%%%%%%%%%%%%%%%%%
%%%%%%%%%%%%%%%%%%%%%%%%%%%%%%%%%%%%%%%%%%%%%%%%%%%%%%%%%%%%%%%%%%%%

\begin{figure*}
\setlength{\unitlength}{1cm}           
\begin{picture}(7,6)       
\put(-6.5,-6.5){\includegraphics{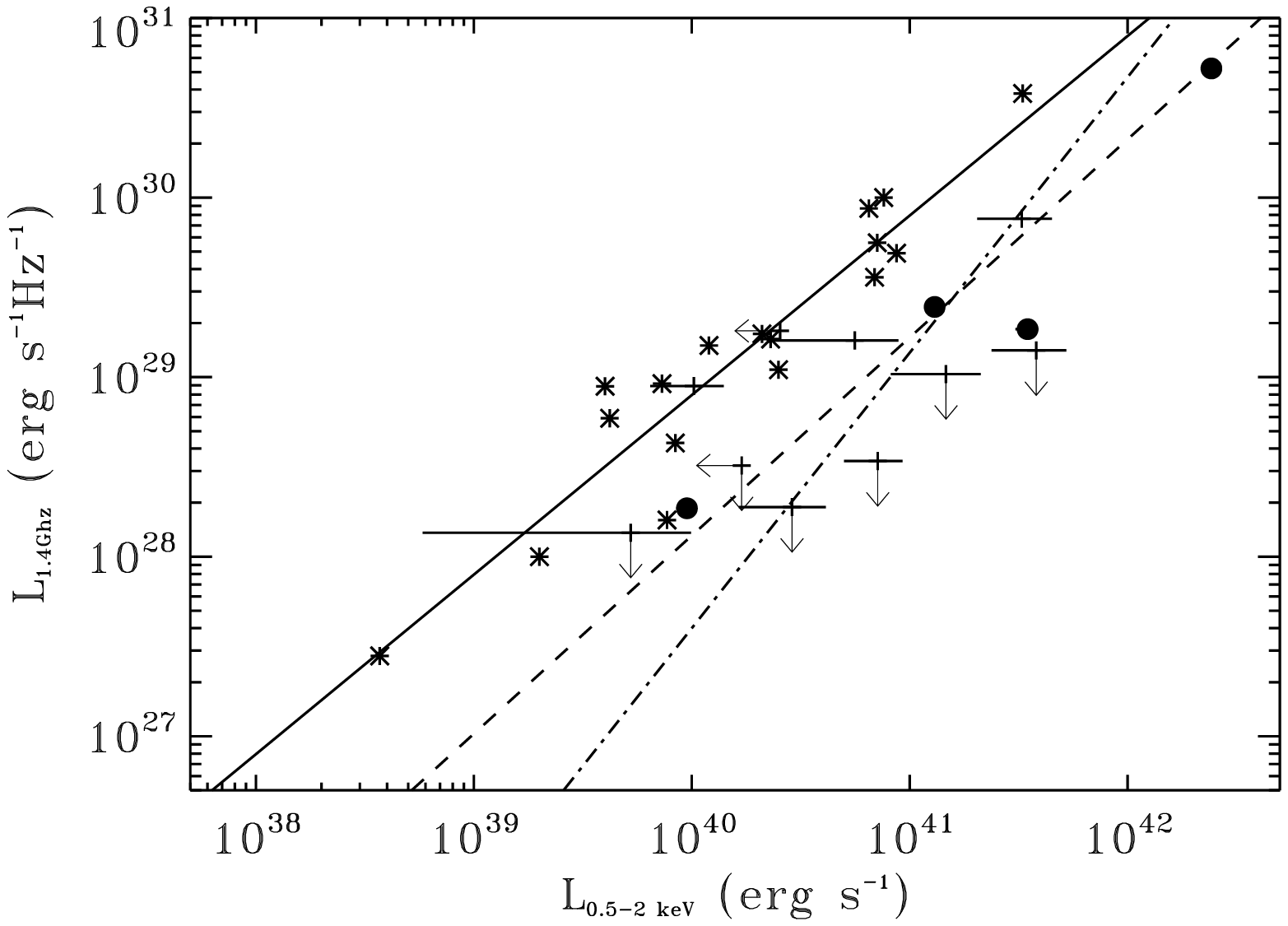}}
\put(2.5,-6.5){\includegraphics{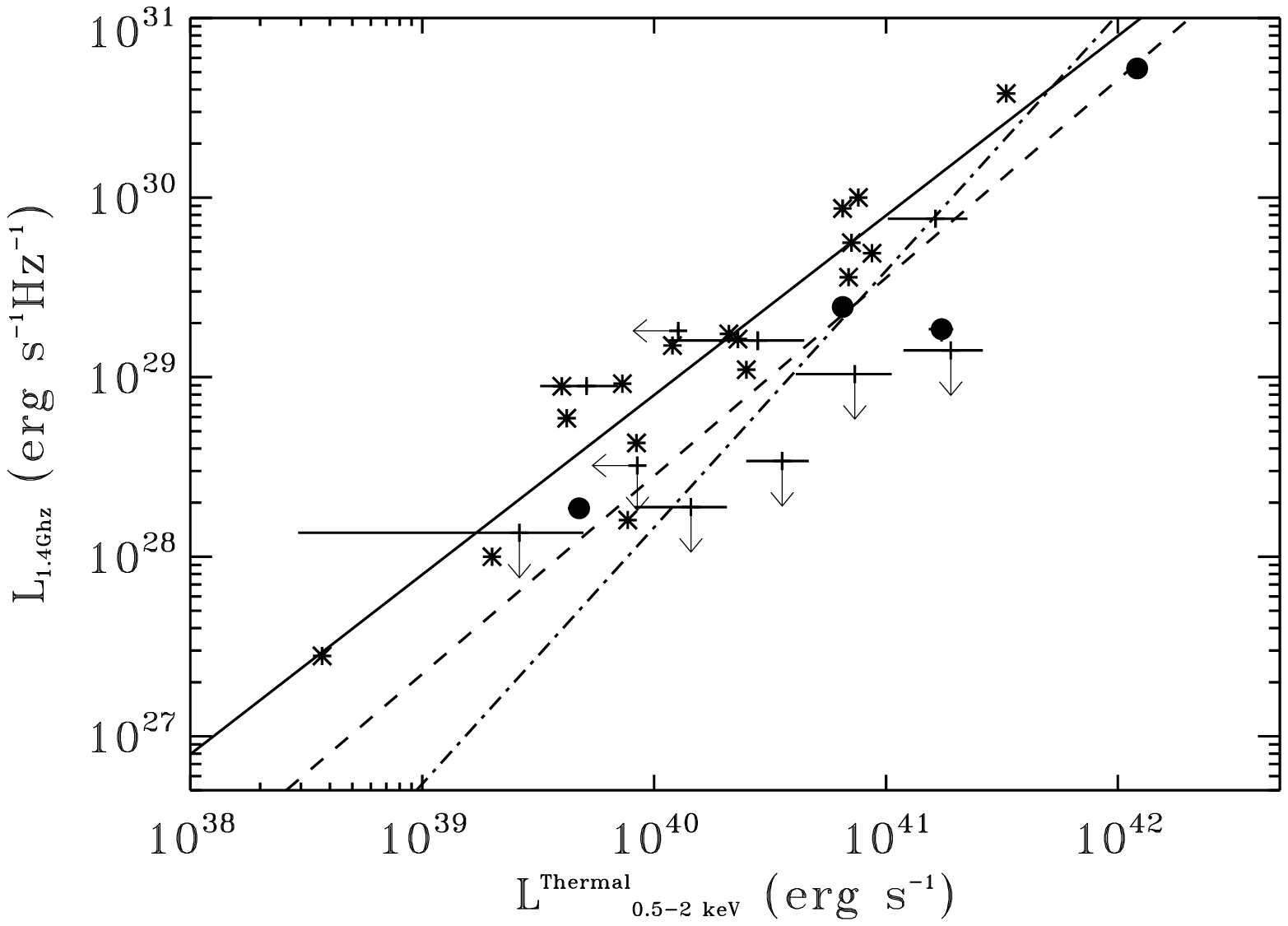}}
\end{picture}
\caption{\label{fig:lxs_lradio} 
Radio  versus soft X-ray luminosity (L$_{0.5-2 kev}$) for the 
selected galaxies.  The X-ray luminosities
have been calculated by using a two model component (thermal + power law) 
in the left panel and  a thermal component in the right panel. 
Galaxies for which it was possible to extract the spectra are marked
with solid circles, the other sources in our sample 
are indicated as crosses.  
In both panels, the galaxies studied by Ranalli et al.
are represented by asterisks. The solid line shows the linear correlation
found by Ranalli and collaborators.
The result of a standard linear fit (dashed line),  
and the fit from the survival analysis (dot---dashed line) are also plotted.
}
\end{figure*}

In Figure~\ref{fig:lxs_lradio}
we compare the radio with the  soft X-ray luminosities together with
the empirical relation given by Ranalli et al.~\cite{2003Ranalli}.
In the left panel the X-ray luminosities
were calculated by using a two component model (thermal + power law
as described in section~\ref{SpecAnalysis}).
Although the original calibration by Ranalli et al.~\cite{2003Ranalli}
is based on the relation between the radio and X-ray luminosities
(solid line), a clear deviation from the empirical relation is obtained for our objects.
Most of the galaxies are below the line defined by Ranalli and collaborators
and seem to have low radio emission. These galaxies could be in an early
stage previous to the synchrotron phase that appear with the first supernova
explosions. At this phase the radio emission is dominated by free--free
electrons produced in HII regions close to massive stars.

To check if a better correlation is found when only the X-ray thermal emission
is considered we use a thermal model to convert the counts to luminosities.
We used PIMMS and a thermal mekal kT=0.8 keV model with solar metallicity  to derive the luminosities 
for all the galaxies in  the sample. 
In this model we assume that we are counting
photons  produced   by  the  interaction  between  the  hot
out-flowing wind  and the ambient gas  in the host galaxy~\cite{2005Grimes}.
In the right panel the X-ray luminosities correspond to the so defined thermal component only.
In both panels the solid line represents the empirical relation of
Ranalli and collaborators and the dashed line, the linear regression to our data.

%%%%%%%%%%%%%%%%%%%%%%%%%%%%%%%%%%%%%%%%%%%%%%%%%

The radio fluxes of 
five of the galaxies  are just upper limits, so we implement the 
BJ algorithm to confirm the observed trends. 
The results are shown by the dot-dashed line in both panels of  figure~\ref{fig:lxs_lradio} 
and in figure~\ref{fig:lxh_lradio}. 

%%%%%%%%%%%%%%%%%%%%%%%%% Figure 9 %%%%%%%%%%%%%%%%%%%%%%%%%%
%%%%%%%%%%%%%%%%%%%%%%%%%%%%%%%%%%%%%%%%%%%%%%%%%%%%%%%%%%%%%%%%%%%%

\subsection{IR and UV luminosities}
As we mentioned before, the IR radiation is a good tracer of the 
SFR averaged over the last 10$^8$ years. 
However an unknown fraction of UV photons could have escaped the
galaxy and consequently, the derived SFR would be underestimated.
To solve this problem Heckman et al.~\cite{1998Heckman} 
and  Buat et al.~\cite{1999Buat} proposed to add the SFR calculated from the ultraviolet to that 
based on the IR. By adding the two contributions, on one hand 
one avoids the complex corrections that must be applied to the
UV data and on the other hand, there is no need to correct for the uncertain 
UV escaping fraction that affects the IR measurements. 
We use the relations~\cite{1998Kennicutt}, 

\begin{equation}
SFR(IR)(M_\odot yr^{-1}) = 4.5\times 10^{-44} L(IR) (erg s^{-1}) 
\end{equation}
\begin{equation}\label{Eq:SFRuv}
SFR(UV)(M_\odot yr^{-1}) = 1.4\times 10^{-28} L_\nu (erg s^{-1}Hz^{-1})
\end{equation}
which are valid for the case of a universal Salpeter initial mass function 
with masses  between 0.1 and 100 \msol. 
The UV data comes from GALEX (MAST Archive) and our UV 
observations made with OM on board \xmm\ (Table~\ref{tab:RadioInfraUV}).
For those galaxies for which IRAS data exist, we estimate both the SFR(IR) 
and SFR(UV). 
%%%%%%%%%%%%%%%%%%%% New Text %%%%%%%%%%%%%%%%%%%%%%%%%%%%%%%%%%%%%%%%
The total SFR for them is  calculated directly by adding the SFR(IR) and 
the SFR(UV) without applying any extinction correction.
%%%%%%%%%%%%%%%%%%%%%%%%%%%%%%%%%%%%%%%%%%%%%%%%%%%%%%%%%%%%%%%%%%%%%%
The results are given in Table~\ref{tab:RadioInfraUV}.
For those galaxies lacking IRAS data, the total SFR is calculated from the 
UV fluxes corrected by extinction, using Equation~\ref{Eq:SFRuv}.
To correct for extinction we use Calzetti's law 
~\cite{2001Calzetti} and the visual extinctions calculated from the
Balmer decrement (Table~\ref{tab:Opt}) but taking into 
account that due to gas surrounding the newly formed stars, they are more 
extinguished than the stars from which most of the UV radiation is coming
[{\it age selective extinction}, e.g. Mayya et al.~\cite{2004Mayya}]. 
Cid~Fernandes et al.~\cite{2005Cid} based on the study of 50,362 galaxies from the Sloan Digital Sky Survey (SDSS) 
found that the relation between 
the visual extinction obtained from the \ha\ to 
\hb\ ratio is $Av= 0.24 + 1.81 Av^{\ast}$, 
where $Av^{\ast}$ is the extinction that is affecting the  continuum light.
The $Av^{\ast}$ values are the ones that we apply to correct the 
observed UV fluxes. 
The comparison of the total SFR with the X-ray luminosities is
presented in Figure~\ref{fig:lxs_TotSFR}.
%%%%%%%%%%%%%%%%%%%% New Text %%%%%%%%%%%%%%%%%%%%%%%%%%%%%%%%%%%%%%%%
The ratio between the soft X-ray and IR luminosities can be used to study the evolutionary
stage of a given star forming burst. The values 
are given in Table~\ref{tab:RadioInfraUV} and a comparison with
the results from recent theoretical models is discussed in the next section.
%%%%%%%%%%%%%%%%%%%%%%%%%%%%%%%%%%%%%%%%%%%%%%%%%%%%%%%%%%%%%%%%%%%%%%

\begin{table*}
\caption{Fluxes and SFRs from radio, infrared, ultraviolet  and total. The
  ratio between the soft X-ray and IR luminosities is shown in the last column. } 
\label{tab:RadioInfraUV}
\begin{tabular}{llrrrrrrrc}
\hline
Name          & F(1.4\,GHz)  &  SFR(1.4\,GHz)  & F(60\,$\mu$m)    
&F(100\,$\mu$m)         &  SFR(IR)  & F(UV)   & SFR(UV)& Tot. SFR &
log$\frac{L_{0.5-2 keV}}{L(IR)} $    \\ 
     \      & (mJy)    & \msol yr$^{-1}$   & (Jy)   & (Jy)     & (\sfr)     & ($\mu$Jy)\ \  & (\sfr) & (\sfr) & \\\hline 
Mrk52         &    13.10$\pm$0.60&   1.10$\pm 0.33$ &4.43$\pm\, 0.03$ &    6.65$\pm\, 0.10$ &    1.47$\pm$ 0.44 &3086.38$\pm36 $  &  0.61$\pm$ 0.31 &   2.09$\pm$  1.2&  -3.51 \\
Cam0902+1448  &    86.84$\pm$0.82&   309$\pm 93$ &4.12$\pm\, 0.21$ &    6.98$\pm\, 0.35$ &   62.1$\pm$ 19.  &1207.58$\pm24 $  & 10.41$\pm$  5.2 &   72.6$\pm$  42&  -2.76 \\
Tol1457-262   &    37.80$\pm$1.80&  14.5$\pm 4.4$ &3.09$\pm\, 0.18$ &    3.68$\pm\, 0.41$ &    4.34$\pm$ 1.3  &1670.91$\pm21 $  &  1.52$\pm$ 0.76 &    5.87$\pm$  3.4&  -2.87 \\
Tol1247-232   &     3.40$\pm$0.50&  10.9$\pm 3.6$ &0.51$\pm\, 0.05$ &$<$ 0.97             &   10.6$\pm$ 3.2  & 943.07$\pm12 $  &  7.17$\pm$  3.6 &   17.8$\pm$ 10. &  -2.83 \\
Tol2259-398   &$<$  1.77         &   2.01           &--               &    --               &    --             & 477.96$\pm3.4$  &  1.29$\pm$ 0.64 &   10.1$\pm$  5.0&  --    \\
Cam08-82A     &     3.06$\pm$0.15&  10.7$\pm 3.2$ &0.58$\pm\, 0.01$ &    0.95$\pm\, 0.16$ &    8.32$\pm$ 2.5  & 203.48$\pm3.9$  &  1.68$\pm$ 0.84 &   10.0$\pm$  5.8&  --    \\
Mrk605        &$<$  1.56         &   1.90$\pm 4.3$ &0.27$\pm\, 0.04$ &$<$ 1.36             &    2.17$\pm$ 0.65 & 172.18$\pm4.3$  &  0.50$\pm$ 0.25 &    2.67$\pm$  1.6&  --    \\
Tol2306-400   &$<$  1.38         &   8.31           &$<$0.25          &$<$ 0.90             & $<$9.86           & 325.57$\pm11 $  &  4.65$\pm$  2.3 &  162$\pm$  81&  --    \\
Mrk930        &    12.20$\pm$0.90&   5.26$\pm 1.6$ &1.25$\pm\, 0.09$ &$<$ 2.15             &    3.52$\pm$ 1.1  & 864.69$\pm18 $  &  0.88$\pm$ 0.44 &    4.40$\pm$  2.6&  -3.89 \\
UM530         &     6.40$\pm$0.60&  45.0$\pm 14$ &0.58$\pm\, 0.07$ &    0.63$\pm\, 0.13$ &   14.5$\pm$ 4.3  & 340.44$\pm17 $  &  5.68$\pm$  2.8 &   20.1$\pm$ 12. &  -2.99 \\
Tol0420-414   &$<$  1.50         &   0.80           &$<$0.25          &$<$ 0.90             & $<$0.87           & 160.44$\pm5.6$  &  0.20$\pm$ 0.10 &    0.20$\pm$  0.1&  --    \\
Tol0619-392   &$<$  1.56         &   6.14           &0.42$\pm\, 0.05$ &$<$ 1.59             &   10.8$\pm$ 3.2  & 238.70$\pm7.8$  &  2.23$\pm$  1.1 &   13.0$\pm$  7.6&  -3.21 \\
UM421         &     4.30$\pm$0.50&   9.46$\pm 3.0$ &0.56$\pm\, 0.07$ &$<$ 1.30             &    8.07$\pm$ 2.4  & 101.48$\pm7.7$  &  0.53$\pm$ 0.26 &    8.60$\pm$  5.0&  -3.51 \\
UM444         &$<$  1.44         &   1.11           &$<$0.25          &$<$ 0.90             & $<$1.27           & 465.66$\pm7.8$  &  0.85$\pm$ 0.42 &    2.28$\pm$  1.1&  --    \\
\hline
\end{tabular}
{\footnotesize  
             }
\end{table*}

\section{Discussion}

The strong hydrogen recombination lines observed in the selected galaxies 
suggest the presence of an intense burst younger than a few million years. 
In most of the cases, the derived SFR 
is higher than that of the archetypal starburst galaxy M82
(Mayya et al.~\cite{2004Mayya}.
In the early phases of the evolution of a star cluster (just before the first
SN explosions) the soft X-rays are mainly produced by the mechanical energy 
released by stellar winds [e.g. Strickland  et al. ~\cite{2002Strickland}, Cervi{\~n}o, Mas-Hesse, 
\& Kunth~\cite{2002Cervino}, Silich, Tenorio-Tagle, \& A{\~n}orve-Zeferino~\cite{2005Silich}], when
the \ha\ luminosity is also stronger (Figure~\ref{fig:lx_lalpha} top panel). 
The hard X-ray luminosity, on the other 
hand, is generally attributed to HMXBs (e.g. Grimm et al.~\cite{2003Grimm} which, at the 
early stages of evolution of these clusters are still in the formation process.
In fact the maximum number of HMXBs occurs 20-50~Myr after the star forming 
event (Shtykovskiy \& Gilfanov~\cite{2007Shtykovskiy}.
Therefore, a lack of hard X-rays is expected with respect to the 
{\it instantaneous} star formation rate traced by the \ha\, luminosity
(Figure~\ref{fig:lx_lalpha} bottom panel).

A dearth of radio emission in the early phases of cluster formation 
has been attributed either to a strong free--free absorption of the 
emitted radio waves (more severe at longer wavelengths), or to an epoch when 
most of the massive stars  have not yet exploded as supernovae [Rosa-Gonz\'alez at al.~\cite{2007Rosa}].
\ha\ and hard X-ray luminosities probably differ because of the time delay 
between the formation of the first SN remnants and the emergence of stellar 
winds in the cluster massive stars.
Therefore it is hard to interpret SFRs  based on different tracers which 
in general are related to different time scale phenomena. 
Figure~\ref{fig:lxs_lradio} (radio vs.~soft X-ray luminosities)
shows the clear signature of a young starburst, where the 
radio luminosity is well below that expected from
the  Ranalli et al.~\cite{2003Ranalli} relation.
We have found that the SFR(1.4\,GHz) is in general lower than the SFR(\ha) 
with the remarkable exception of Cam~0902+1448 for which the 
SFR(1.4\,GHz) is about 6 times larger than  the SFR(\ha).
When we compare the radio luminosities with the soft X-ray luminosities and 
the  Ranalli empirical relation
the deficit in radio is noticeable (Figure \ref{fig:lxs_lradio}). 
However, when we compare the  radio luminosities 
with the fraction of the  soft X-ray luminosities due to thermal 
emission alone, a better agreement is found (see right panel of 
Figure~\ref{fig:lxs_lradio}). Notice that when the BJ method is used to 
estimate the linear regression, a  higher discrepancy is observed.

\begin{figure}
\setlength{\unitlength}{1cm}           
\begin{picture}(7,6)       
\put(-1.,-6.5){\includegraphics{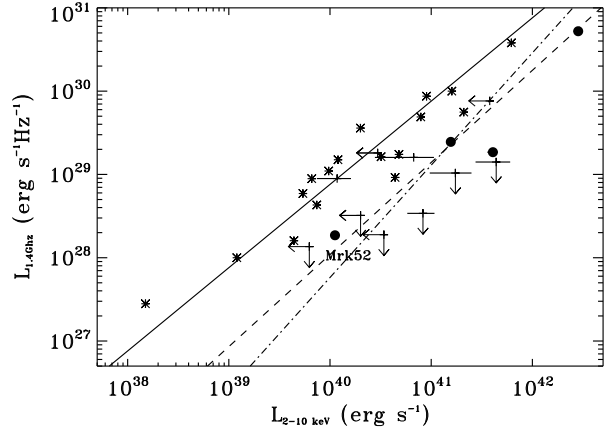}}
\end{picture}
\caption{\label{fig:lxh_lradio} Radio vs. hard X-ray luminosities. Symbols as in
Figure \ref{fig:lxs_lradio}.
}
\end{figure}

Radio  versus hard X-ray luminosities  are shown in 
Figure~\ref{fig:lxh_lradio}. We can see that most of the galaxies are below 
the relation proposed by Ranalli et al.~\cite{2003Ranalli}, consistent with 
the lack of both SN remnants and HMXBs.

\begin{figure}
\setlength{\unitlength}{1cm}           
\begin{picture}(7,6)       
\put(-1.5,-7.25){\includegraphics{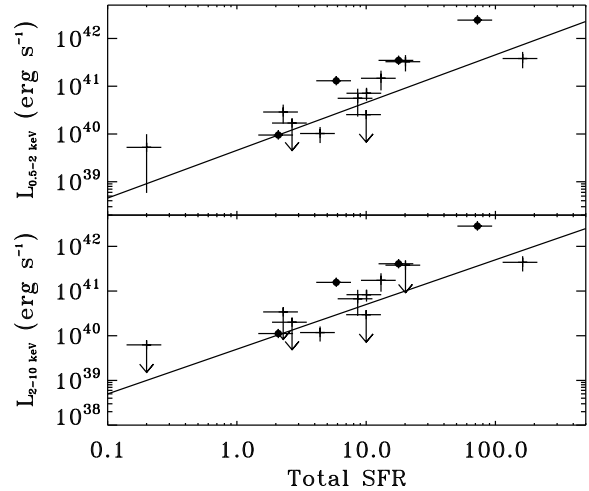}}
\end{picture}
\caption{\label{fig:lxs_TotSFR} 
X-ray luminosity (top panel for the soft and bottom panel for the
hard band) against the total SFR. 
The line shows the Ranalli empirical law for soft and hard X-ray luminosities.
In both panels, solid circles are the bright X-ray objects in the sample with an
X-ray spectrum. 
}
\end{figure}

For galaxies with avaliable IRAS data, the total SFR was calculated  by adding the SFR(IR) and the
SFR(UV) without correcting the UV fluxes for extinction.
For those galaxies without IRAS fluxes, 
the total SFR was calculated from the extinction corrected UV luminosities.
The total SFR is an average over the last 10$^8$ years, providing enough time 
for a population of HMXB to be in place. 
In fact there is a good correlation 
between the total SFR (averaged over the last 10$^8$ years) and the X-ray 
luminosities both in the soft and in the hard band 
(Figure~\ref{fig:lxs_TotSFR}), in fairly good agreement with the empirical 
Ranalli relation.

%%%%%%%%%%%%%%%%%%%%%%%%%%%%%%%%%%%%%%%%%%%%%%%%%%%%%%%%%%%%%%%
%%%%%%%%%%%%  New text   %%%%%%%%%%%%%%%%%%%%%%%%%%%%%%%%%%%%%%
Notice that all selected galaxies are classified as HII galaxies based on
BPT diagrams of optical line ratios. However an unknown contribution from a heavily
obscured AGN could be present. This contribution is more important in
the hard band,
and the deviation from the Ranalli relation observed in
Figure~\ref{fig:lxh_lradio}
could be due to an excess of X-ray emission from an AGN
instead of representing a deficit in radio. However, such deficit in radio is clear
when the SFR(H$\alpha$) is compared with the SFR(1.4 GHz). Moreover,
the excess of
hard X-ray emission is not present when comparing the hard band X-ray luminosity with the total 
SFR (Figure~\ref{fig:lxs_TotSFR}).
We can safely conclude that the X-ray emission that we see is due to
star formation processes.

%%%%%%%%%%%%%%%%%%%%%%%%%%%%%%%%%%%%%%%%%%%%%%%%%%%%%%%%%%%%%%%
%%%%%%%%%%%%%%%%%%%%%%%%%%%%%%%%%%%%%%%%%%%%%%%%%%%%%%%%%%%%%%%

%%%%%%%%%%  Comparison with MM theoretical results %%%%%%%%%%%%
%%%%%%%%%%%%%%%%%%%%%%%%%%%%%%%%%%%%%%%%%%%%%%%%%%%%%%%%%%%%%%%
Mas-Hesse et al.~\cite{2008Mas} have concluded that assuming a reprocessing efficiency of the
mechanical energy of a few percent, the observed values of the ratio between 
the soft X-ray and infrared luminosities can be reproduced with synthetic 
models of young starbursts. In fact they obtain an average value of 
$log\,L_{0.5-2keV}/L_{IR}= -3.5$
for  efficiencies between 1\% and 10\% for bursts with 
ages around 5 Myr.
In our sample,  the ratio $log\,L_{0.5-2keV}/L_{IR}$ (Table~\ref{tab:RadioInfraUV}) 
has an average value of $-3.20\pm 0.41$ consistent with the theoretical
estimations. 
%%%%%%%%%%%%%%%%%%%%%%%%%%%%%%%%%%%%%%%%%%%%%%%%%%%%%%%%%%%%%%%
%%%%%%%%%%%%%%%%%%%%%%%%%%%%%%%%%%%%%%%%%%%%%%%%%%%%%%%%%%%%%%%
%%%%%%%%%%%%%%%%%  New text %%%%%%%%%%%%%%%%%%%%%%%%%%%%%%%%%%%
%%%%%%%%%%%%%%%%%%%%%%%%%%%%%%%%%%%%%%%%%%%%%%%%%%%%%%%%%%%%%%%
For continuous star formation,
a value of $log\,L_{0.5-2keV}/L_{IR}= -3.5$ is consistent
with ages as large as 25~Myr. At this age,  a significant number of
stars have exploded as SN. The strong emission lines
and the high SFR derived together with the
lack of synchrotron emission observed in most of the sources
suggest a star formation history characterized
by a strong short burst.
%%%%%%%%%%%%%%%%%%%%%%%%%%%%%%%%%%%%%%%%%%%%%%%%%%%%%%%%%%%%%%%
%%%%%%%%%%%%%%%%%%%%%%%%%%%%%%%%%%%%%%%%%%%%%%%%%%%%%%%%%%%%%%%

Mrk~52 and Cam0902 are outliers in some of the  presented plots. 
Particularly, Mrk~52 is a bright blue galaxy with a derived SFR(\ha)=14~\sfr.
However, the extremely weak emission in the hard X-ray band  
(see Figure~\ref{fig:mrk52}) together  with the deficit in 
radio makes this galaxy the most clear example of an extreme young burst with 
an age of about 3~Myrs when the majority of supernovae and  HMXBs have not yet 
been formed. This age is consistent with Mrk~52 being a Wolf-Rayet galaxy as 
the dominant burst is going through an intense Wolf-Rayet phase 
[Schaerer, Contini, 
\& Pindao~\cite{1999Schaerer}, Fernandes et al.~\cite{2004Fernandes}].

%%%%%%%%%%%%%%%%%%%%%%%%% Figure 10 %%%%%%%%%%%%%%%%%%%%%%%%%%
%%%%%%%%%%%%%%%%%%%%%%%%%%%%%%%%%%%%%%%%%%%%%%%%%%%%%%%%%%%%%%%%%%%%

%%%%%%%%%%%%%%%%%%%%%%%%% Figure 11 %%%%%%%%%%%%%%%%%%%%%%%%%%
%%%%%%%%%%%%%%%%%%%%%%%%%%%%%%%%%%%%%%%%%%%%%%%%%%%%%%%%%%%%%%%%%%%%

%
%______________________________________________________________

\section{Conclusions}
We have analyzed \xmm\ observations of a sample of 14 star forming galaxies with high
SFR judging from their strong hydrogen recombination lines. 
The main conclusions of our analysis are: 
\begin{enumerate}
\item The SFR obtained from the soft X-ray luminosity is comparable to that 
determined from the \ha\ luminosities. This is consistent with the X-ray 
luminosity as a good tracer 
of SFR even at early times when the massive stars dominate 
the energy generation. Due to the time delay between 
the formation of the burst and the formation of the first 
massive binary systems, a lack of hard X-ray luminosity is observed
when compared with tracers of recent star formation activity (e.g. \ha, Figure~\ref{fig:lx_lalpha}).
This effect has already  been observed in the Small Magellanic Cloud;  we have
extended the study to a larger sample of starforming galaxies.  

\item The weak radio emission observed in some of the objects 
(related to early phases in the starburst evolution)
together with the values found for $log\,L_{0.5-2keV}/L_{IR}\sim$ --3.2
suggests that the sample of galaxies is biased to galaxies dominated by 
young bursts with ages lower than 5 Myrs and with star forming histories 
characterized by long and low activity periods, followed by strong short episodes.
\item The relation of both soft and hard X-ray luminosity with
the SFR traced by the IR and UV, shows that the X-ray production 
is maintained during at least 10$^8$ years. 
\end{enumerate}

\section*{Acknowledgments}
D. R. G. and E. J. B.  thank the hospitality of  ESAC (Madrid) where most of the
reduction and analysis of the \xmm\ data was realized. 
The visit was possible thanks to an ESAC Faculty grant.
We thank Miguel Mas Hesse, H\'ector Ot{\'{\i}} Floranes and Divakara Mayya for useful discussions.
D. R. G., R. T., and E. T. acknowledge
support by the Mexican Research Council (CONACYT) under
grants 49942 and 40018.

The authors are very grateful to an anonymous referee whose
comments and suggestions largely improved the clarity of this paper.

\bsp
\label{lastpage}
\end{document}